# Potential Energy Landscape as a Framework for Developing Innovative Materials

Nadezhda A. Andreeva and Vitaly V. Chaban

(1) Peter the Great St. Petersburg Polytechnic University, Saint Petersburg, Russia. E-mail: andreeva_na@spbstu.ru

(2) Yerevan State University, Yerevan, 0025, Armenia. E-mail: vvchaban@gmail.com

**Abstract.** In the contemporary era of rapid advancements in materials science, the development of new compounds and materials is proceeding at an accelerated pace. The concept of the potential energy landscape (PEL) plays a pivotal role in supporting the meticulous engineering of novel structures. This review article examines the historical evolution of the PEL concept and its diverse applications in materials design. A comprehensive overview of the major methods employed to sample the PEL is presented, accompanied by a critical discussion highlighting relevant modern endeavors. Specific applications of the PEL in rationalizing the design of molecules and materials for energy storage, electrolytic solutions, and greenhouse gas capture are exemplified. This review serves as an up-to-date guide for exploring and analyzing the PEL, facilitating a deeper understanding of its significance in materials science.

**Keywords:** potential energy landscape; materials design; electrolyte; metalized film capacitor; $CO_2$ scavenger.

# 1. Introduction

The field of materials science is currently experiencing a period of rapid growth, with thousands of novel products having been developed and commercialized in recent decades. This progress in applied sciences leverages the theoretical foundations established in physics, chemistry, and biology. Generalized concepts and models serve to rationalize and support modern developments and their practical applications.

In this context, the potential energy landscape (PEL) emerges as a foundational concept in materials science.[1-3] The PEL provides a robust and convenient framework for comprehending material behavior at the atomic and molecular levels. It offers a visual and conceptual representation of a system's potential energy as a function of its coordinates. The system under investigation can comprise atoms, molecules, ions, or even model particles such as coarse grains. By examining the topography of the PEL, researchers can gain comprehensive insights into the thermodynamic stability, dynamics, and phase transitions of materials. Each material, whether pure or composite, possesses a unique PEL.[4-6]

Mathematically, the PEL is a multi-dimensional surface representing the potential energy of a system as a function of the configuration of its constituents (atoms, molecules, ions). The configuration can be defined by various parameters, such as the positions of interaction centers, bond lengths, valence angles, and other geometry-related descriptors. The topography of the PEL, with its valleys, peaks, and saddle points, governs the properties and behavior of the material or sample under consideration. The PEL is an extensive physical property, meaning its size is directly proportional to the sample size, regardless of whether the sample structure is regular or amorphous. In the case of non-doped crystals, it is reasonable to consider only the PEL of the elementary lattice. Due to its multi-dimensional nature, the entire PEL cannot be conventionally visualized. However, it can be analyzed using methods of differential analysis.[7-9]

## 2. Theoretical Background

The PEL concept is rooted in the development of statistical mechanics and thermodynamics. Pioneers in statistical mechanics during the late 19th and early 20th centuries, such as Josiah Willard Gibbs and Ludwig Boltzmann, laid the foundation for understanding the relationship between energy and the microscopic configurations of a system.[10] The PEL framework emerged as a natural extension of these ideas, providing a visual representation and convenient terminology for comparing and discussing the energy landscapes of various thermodynamic phases and materials (Figure 1).

Gibbs introduced the concept of ensembles, which are collections of a large number of identical systems, to describe the macroscopic behavior of matter based on microscopic principles. The idea of ensembles contributed to understanding how the energy distribution within a system relates to its observable properties. Boltzmann further advanced statistical mechanics by linking entropy, a measure of disorder, to the number of possible microscopic states of a system. This principle is now known as the Third Law of Thermodynamics,

$$S = k \ln W.$$

In this equation, $S$ represents entropy, $k$ is Boltzmann's constant, and $W$ denotes the number of microstates. This equation established a crucial link between the microscopic and macroscopic descriptions of matter.

Maxwell's development of the kinetic theory of gases provided a statistical description of how individual gas molecules, with their varying speeds and collisions, give rise to macroscopic properties such as pressure and temperature. The kinetic theory further solidified the importance of statistical approaches in understanding matter. The Boltzmann distribution of momenta links them to the potential energies of the interacting particles and the temperature of the system,

$$p_i \propto e^{\frac{-\varepsilon_i}{kT}},$$

where $p_i$ is the probability of the system being in state i, $\varepsilon_i$ is the energy of that state, and a constant kT of the distribution is the product of the Boltzmann constant k and thermodynamic temperature T. The distribution describes the probability of finding a system in a particular energy state depending on its potential energy. Furthermore, it reveals that lower-energy microscopic states are more probable within the macroscopic state of matter. In other words, the abundance of low-energy states stems from their stability. It is important to note that the role of low-energy states increases exponentially as their potential energy decreases. In most conventionally defined physicochemical systems, the potential energies of the most abundant microscopic states are negative, while the thermal motion-induced kinetic energy is positive.

In the context of the PEL, the theoretical advances formulated by Maxwell and Boltzmann imply that systems tend to occupy the valleys of the landscape. These potential energy valleys correspond to the most stable molecular, atomic, or ionic configurations. Early work on chemical reactions and molecular structures began to visualize the relationship between potential energy values and microscopic configurations. These advances led to the emergence of concepts such as the reaction coordinate and potential energy surface, which are essentially lower-dimensional representations of the PEL.[11-15]

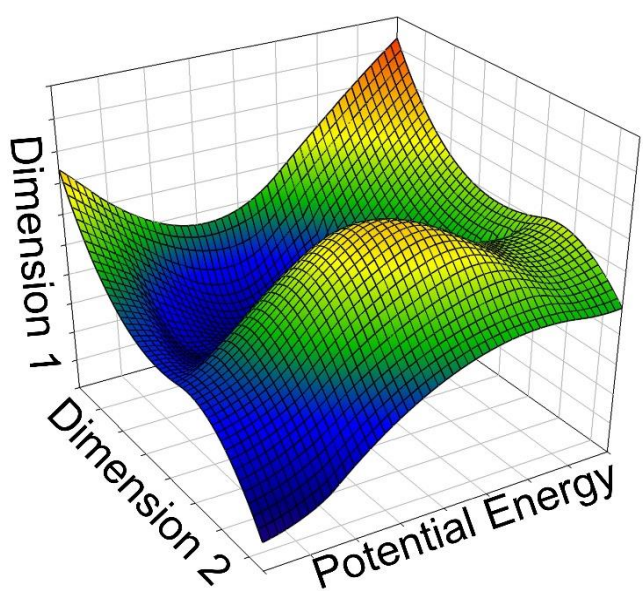

Figure 1. Exemplary potential energy landscape of a simplified model system.

## 3. Evolution and Current State of the PEL Concept

The rise of computational power, beginning in the latter part of the 20th century, enabled the application of the PEL concept in materials science. As computers gained power, researchers could perform complex calculations and simulations of atomic and molecular systems, facilitating detailed exploration of the PEL and its connection to material properties. This progress necessitated the development of specific theoretical models to represent interacting microscopic objects. Quantum chemical calculations required models to overcome the limitations of the wave equation for many-atom systems, while classical simulations relied entirely on newly derived models. Advances in quantum mechanics and molecular dynamics (MD) simulations provided the tools to calculate the potential energy of complex systems, leading to a refined understanding of PEL features.[16-18]

Martin Goldstein coined the term "energy landscape" in 1969 to describe the complex potential energy surface of glasses and the numerous metastable states they can occupy.[19] The PEL framework continued to evolve alongside new theoretical approaches and computational techniques, providing a powerful lens for understanding material structure, properties, and behavior. Today, the PEL bridges the gap between microscopic systems of atoms and molecules and the macroscopic materials we study.

The modern definition of the PEL has evolved since its introduction. Early depictions focused on simple 2D or 3D representations, but modern definitions recognize the incredible complexity and high dimensionality of PELs for real-world materials. Each atom or molecule contributes degrees of freedom, making direct visualization impossible. The PEL is now frequently described mathematically as a function of many variables, each representing a coordinate in the system's configurational space. Computer simulations can parameterize the PEL using specific constants once a model Hamiltonian is assigned.

While the PEL can be visualized as a static landscape, current definitions emphasize its dynamic nature, particularly for systems sensitive to thermal motion. Atoms constantly explore their PELs through vibrations, rotations, and diffusion. Researchers are increasingly interested in how PELs change over time, especially during chemical reactions and phase transitions. Accurately capturing PEL features requires considering the complex interplay of interactions, including many-body effects. A quantum chemical sampling of potential energy variations is preferable despite being computationally expensive compared to classical pairwise force fields. For materials with light elements or at low temperatures, quantum mechanical effects necessitate more sophisticated models.

The PEL of a material can be parameterized using in silico simulation methods. Density functional theory (DFT) is one efficient approach for evaluating potential energy values at different points on the landscape, whether specified by researchers or determined algorithmically. DFT also provides electronic structures and spectroscopic properties. Machine learning is another approach, identifying patterns in PEL data and predicting new materials by finding hidden regularities. Multiscale modeling bridges the gap between atomic-level PEL and macroscopic behavior.[20-21]

A high-quality parameterized PEL must be connected to real-world observations. Modern PEL research emphasizes correlating theoretical predictions with experimental evidence. Techniques like atomic force microscopy, X-ray diffraction, and neutron scattering provide detailed information about atomic structure and dynamics, allowing for direct probing of the PEL.[22-24]

The PEL has become an indispensable tool in materials science, enabling researchers to predict and understand material properties, design new materials, compare similar materials, and optimize processing techniques. PEL characteristics vary significantly depending on the material. Here are a few examples highlighting various types of chemical structures:

Crystalline solids are characterized by a highly ordered arrangement of atoms or molecules. Their PELs exhibit well-defined energy minima corresponding to the stable crystal structure. Modifications, like atomic substitutions, alter the PEL with perturbation magnitudes proportional to the structural differences. Defects and dislocations introduce local distortions in the PEL, influencing mechanical properties and thermodynamic stability. Reinforcing agents positively impact material properties, which is reflected in the PEL. Phase transitions induced by temperature or pressure changes are represented by changes in global and local energy minima on the PEL.[11, 25-26]

Amorphous solids lack the long-range order of crystalline materials. Their PELs are complex, multi-dimensional surfaces with numerous local energy minima and transition points. The precise shape depends on thermodynamic conditions. The glass transition, a gradual transition from a liquid-like to a solid-like state, is well-described by the PEL framework. Polymers have PELs influenced by chain length, flexibility, and intermolecular interactions. The PEL aids in understanding the mechanical and thermal properties of both glasses and polymers.[27-29]

Liquids are characterized by fluidity. Their PELs are dynamic surfaces with constant energy minima due to thermal motion. Liquids exhibit shallow energy minima, allowing for easy molecular rearrangement. The PEL's dynamic nature is responsible for liquid fluidity. Nucleation and crystallization from liquids are described within the PEL framework, which helps predict energy barriers and link them to thermodynamic conditions.[30-32]

## 4. Applications of the PEL Framework

The PEL framework has found widespread applications in materials physics and chemistry, particularly in understanding the nuanced properties of similar materials, predictive modeling and simulations, high-throughput screening, and optimization. Computational techniques such as MD, Monte Carlo (MC) simulations, stationary point searches, and their combinations provide algorithms to explore the PEL and predict material properties by moving atoms and ions through phase space. This enables the screening of a vast number of potential materials to identify promising candidates with desired physicochemical and electrical properties for innovative applications. The PEL framework also elucidates how materials respond to external forces, including elasticity, plasticity, and fracture, and predicts thermal properties like heat capacity and thermal conductivity. Furthermore, it aids in understanding the evolution of electronic structures and predicting electrical conductivities.

Annealing, a crucial practice in both real-world processes and computer modeling, aims to achieve the most thermodynamically stable conformations of investigated species. This technique is particularly beneficial for chemical structures with complex PELs. Annealing involves heating a material to a high temperature and then slowly cooling it, with the specific temperatures, known as the annealing regime, selected based on the material's phase behavior, which can be hypothesized from the modeled PEL. The PEL framework helps understand how thermodynamic conditions affect material microstructure and properties.[33-34]

The PEL framework also optimizes deposition from solutions and crystal growth techniques, such as thin film deposition and the production of crystals with desired shapes, properties, and structural features. The PEL concept supports cutting-edge developments in materials science, where potential energy variations as a function of coordinates are crucial.

The PEL framework continues to evolve with advancements in mathematical and experimental techniques. Machine learning algorithms analyze PEL data to identify unusual patterns, while experimental techniques like atomic force microscopy and X-ray diffraction provide insights into the PEL at the atomic level of resolution. Multiscale modeling bridges the gap between atomic and macroscopic scales, providing a more complete understanding of material behavior.

The PEL serves as a guiding principle in materials science, offering a powerful tool for understanding and predicting material behavior across various classes and properties. Its topography, with valleys, peaks, and saddle points, reflects these properties and behaviors.

Crystalline solids, with their highly ordered atomic arrangements, correspond to the lowest possible enthalpy but also the lowest possible entropy due to the lack of conformational freedom in the lattice. This translates to a PEL with deep, well-defined energy minima, indicating a stable crystal structure where atoms exist in their lowest energy configuration, the global minimum. The PELs of crystals reflect their periodic patterns, with repeating potential minima extending throughout the material.

Defects and dislocations in crystals significantly alter the PEL compared to the ideal structure. Natural crystals inevitably have imperfections, such as vacancies, interstitials, or misaligned planes of atoms, corresponding to local minimum stationary points in the PEL. While thermodynamically less stable than the ideal lattice, many distorted crystal structures remain stable for extended periods due to being surrounded by high-energy barriers in the PEL. These distortions disrupt the PEL's periodicity, creating local "bumps" or "hills" representing higher potential energy regions around defects. Shallow minima near dislocations allow for more effortless movement of atoms along the dislocation line, influencing material properties like ductility.

Phase transitions and transformations are readily discussed and quantified using the PEL. Melting, freezing, and changes between different crystal structures involve shifts in the global and local energy minima of the PEL. Temperature and pressure can alter the PEL, affecting the prominence of certain minima.

As the glass cools, atomic and molecular movements slow down dramatically, and the system can get trapped in one of the many local minima on the PEL, even if lower energy states exist. The glass transition is a gradual change from a liquid-like to a solid-like state, and amorphous solids exhibit a disordered landscape without long-range order. Their PELs are complex and "rugged," featuring numerous local energy minima but no repeating pattern. This "frustration" can trap atoms in deep local minima, preventing them from reaching the actual lowest energy state.

Polymers exhibit unique PELs due to their long, flexible molecular chains and interchain couplings. Each possible conformation of the polymer chains corresponds to a local minimum in the PEL, and attractions and repulsions between chain segments further complicate the PEL. The energy landscape influences the flexibility, elasticity, and other properties of polymers.[35-37]

Liquids share some features with polymeric systems. Atoms or molecules in liquids constantly move and rearrange, resulting in shifting and changing local minimum points on the PEL. Liquids are characterized by shallow minima, allowing for easy movement of atoms and explaining their fluidity. The average distance between local energy minima determines particle density.

Activation energy barriers are understood by analyzing the PEL shape. For example, crystallization from a liquid requires overcoming an energy barrier to reach a lower energy crystalline state. The shape of the PEL and the energy difference between liquid and crystalline states determine the energetic cost of the process. The stability of supersaturated and superheated solutions can be fully characterized by investigating their PELs.

In the context of mechanical properties, deep and narrow energy valleys suggest strong bonds and resistance to deformation, indicating a brittle material, while shallower and wider valleys suggest weaker bonds and greater ease of atomic rearrangement, indicating a more ductile material. Barriers separating energy valleys represent the potential energy needed for dislocations to move, influencing strength and plasticity.

For thermal properties, the PEL predicts thermal conductivity and heat capacity. The steepness of the PEL affects how easily atoms vibrate and transmit heat, and the density of local minima influences how much energy a material can store.

Regarding phase behavior, the PEL helps understand and predict phase transitions like melting, boiling, and changes in the crystal structure, which correspond to transitions between different energy minima.[38-39]

Computational exploration of the PEL involves using techniques like MD, MC, and geometry optimizations to explore the PEL of hypothetical materials before actual preparation. This allows for property prediction and identification of promising candidates with desired characteristics. By understanding how atomic arrangements and compositions affect the PEL, scientists design materials with specific properties, such as high strength, lightweight, or tailored electronic behavior. The PEL framework enables high-throughput screening of vast numbers of potential materials in silico, accelerating the discovery process and reducing reliance on costly experimentation.

In summary, the PEL provides a unifying concept for understanding the relationship between atomic-scale structure and macroscopic properties. By navigating this landscape, scientists gain a deeper understanding of existing materials and pave the way for the design and discovery of new materials with revolutionarily fine-tuned properties. The examples provided for different classes of atomic and molecular systems illustrate how the PEL framework helps us understand the diverse properties and behaviors of materials.

## 5. Practical Real-World Calculations of the PEL

The inherent multi-dimensionality of the PEL presents challenges for its full parameterization in systems of practical size. While parameterizing PELs for simple systems like water dimers in vacuo, real gases, or solids with high degrees of symmetry is feasible, systems of practical interest often possess too many dimensions to explore systematically. However, full parameterization is usually unnecessary when addressing real-world challenges in materials design.

For instance, optimizing a chemical reaction requires parameterizing only a few bond dimensions related to bond breakage and formation, while other PEL dimensions remain less critical. Developing an optimized electrolyte system necessitates a quantitative understanding of electrode-electrolyte and electrolyte-electrolyte electrostatic couplings to estimate the properties of the energy storage device (Figure 2).[40] Even in materials design endeavors requiring the characterization of a more significant fraction of the PEL, information about absolutely all stationary points is not essential.

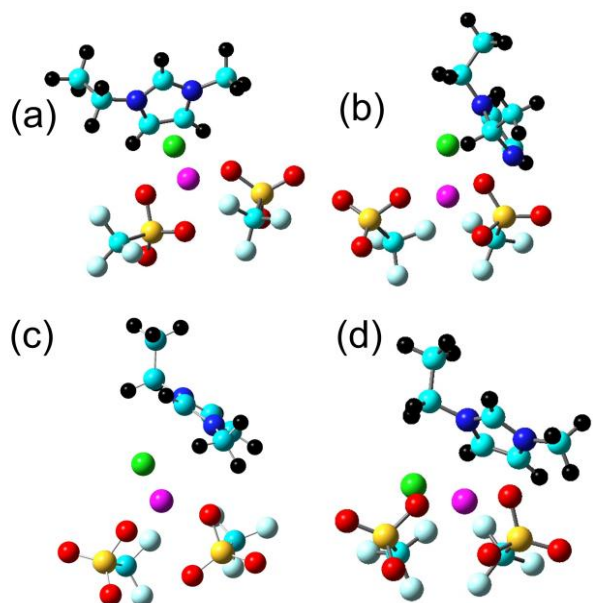

Figure 2. (a) The global minimum geometry of the [Mg][OTf]$_2$ + [EMIM][Cl] system with a potential energy of -2873 kJ/mol; (b) the local minimum geometry with a potential energy of -2671 kJ/mol; (c) the local minimum with a potential energy of -2848 kJ/mol; (d) the local minimum with a potential energy of -2705 kJ/mol. All enumerated potential energies correspond to the implicitly simulated DME. The carbon atoms are cyan, the hydrogen atoms are black, the nitrogen atoms are blue, the fluorine atoms are light blue, the oxygen atoms are red, the sulfur atoms are yellow, the chloride anions are green, and the magnesium cations are violet. Reproduced from Ref.[40] with permission from the Elsevier.

High-energy stationary points are irrelevant due to their limited stability in a real functional material. Although high-energy local minima represent valleys of thermodynamic stability, their lifetimes are short, as they quickly decay to more stable structures through thermal motion. Most chemical transformations are powered by rare events that are difficult to observe in real-time MD simulations, lasting picoseconds to microseconds yet occurring over macroscopic periods.

Medium-energy stationary points are accessible only at significantly elevated temperatures. If the investigated system exhibits low thermal stability, further consideration of such minima is unnecessary. The lowest-energy saddle point corresponding to the onset of material decomposition determines structural evolutions. The initially parameterized PEL contains all these energies, making it crucial to begin characterization by identifying all stationary points and ranking their energies. Depending on the research goals, internal energy, enthalpies, Helmholtz free energy, or Gibbs free energy may be the most relevant thermodynamic potentials for assessing the importance of the identified stationary points. This initial identification can be conducted using computationally inexpensive model Hamiltonians, such as semiempirical or tight-binding DFT methods.

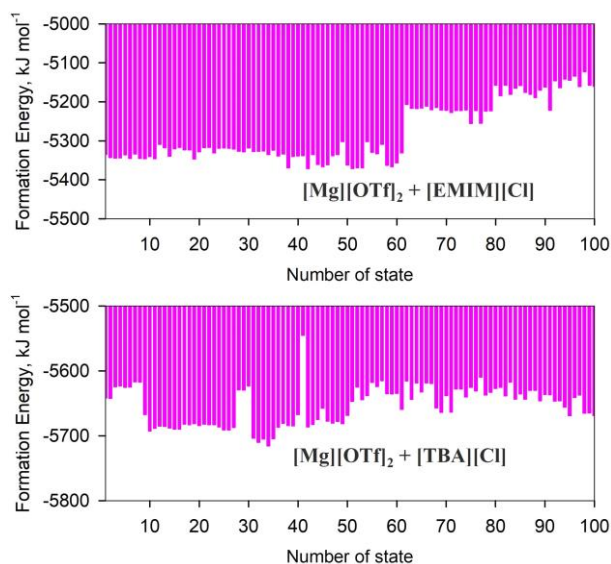

Figure 3. The distributions of the stationary point energies in $[Mg][OTf]_2$ + $[EMIM][Cl]$ and $[Mg][OTf]_2$ + $[TBA][Cl]$ electrolyte solutions. Six molecules of DME represented the solvation effects. Reproduced from Ref.[40] with permission from the Elsevier.

Typically, only a few low-energy local minima, along with the global minimum, play essential roles in determining the macroscopic system's behavior. As a rule of thumb, energy differences of a few k×T products from the global minimum stationary point should be identified and characterized.

Only the most low-energy transition states corresponding to key chemical transformations are of interest. These transition states must be simulated, and their potential energies, including zero-point energy corrections, must be compared. Identifying transition states is more complicated, especially for arbitrary PELs, as it requires specific numerical methods to move uphill along the reaction coordinates.[41]

In summary, parameterizing the entire PEL for every material of technological interest is unnecessary. Instead, characterizing a set of low-energy minima and saddle points provides information about the stability, dominant structures, and properties of the investigated species. If the material's properties and behaviors are known, specific PEL dimensions can be chosen to derive desired descriptors, such as reactivities with chemical agents or the microscopic states exhibiting the most promising electronic and thermal conductivities.[42]

## 6. Modern Advances in Unraveling the PEL

Machine learning is being employed to accelerate the exploration and enhance understanding of the PEL's complexities. Brute-force methods for mapping PELs are computationally expensive, but machine learning algorithms can potentially expedite this process by learning the relationships between atomic configurations and their corresponding potential energies, enabling efficient exploration of vast and complex PELs. This approach effectively substitutes the task of solving wavefunctions by identifying structurally similar patterns and assigning appropriate potential energies.[43-45]

Currently, machine learning for PELs is in its early stages, and the accuracy of automated energy assignments based on known structural regularities remains to be fully established. It is more appropriate to consider machine learning algorithms as preliminary tools for identifying key PEL dimensions and suggesting a rough overall shape. Such calculations are not expected to require supercomputing resources and can be performed quickly.[46-47]

Machine learning can predict material properties from associated PELs. Trained models can predict properties like melting and boiling points, stability, and conductivity directly from the PEL, bypassing expensive and time-consuming simulations and experiments. Similar molecular patterns often correspond to comparable properties, and machine learning models can identify non-intuitive regularities. By quickly exploring various types of chemical thinking, machine learning uncovers hidden patterns and correlations within PELs, leading to new insights into the relationship between structure and properties and facilitating the fine-tuning of existing materials.[48-49]

By learning the characteristics of stable and low-energy configurations, machine learning can generate new materials with slightly altered physicochemical and electrical properties by exploiting hidden structure-property relationships. Importantly, machine learning does not replace computational chemistry methods but rather simplifies the process of finding sophisticated research schedules with minimal human intervention.

Experimental techniques, such as X-ray diffraction, spectroscopies, and atomic microscopies, are essential for validating machine learning model predictions about PELs and material properties. While deriving potential energies directly from experiments is challenging, machine learning benefits from experimental data, which can train and refine models, improving accuracy and predictive power. This creates a synergistic loop where experiments inform machine learning, and machine learning guides further experiments. Techniques like ultrafast spectroscopy probe the dynamic evolution of the PEL in real-time. They provide valuable data for training machine learning models to understand and predict material behavior under non-equilibrium conditions, such as during chemical transformations.

Multiscale modeling and simulation using machine learning and the PEL concept are of central research interest. Since material properties emerge from microscopic organization at different spatial scales, multiscale modeling is crucial. Machine learning integrates with multiscale modeling approaches to connect atomic-scale information from conventional PELs with coarse-grained PELs. For example, many biological systems exhibit separate behaviors at different scales, requiring simulations at both atomistic and microscale levels. Multiscale data can be connected to macroscopic properties, leading to a more complete understanding of how microscopic features influence macroscopic behaviors. Machine learning-directed development of coarse-grained PEL representations reduces computational cost while preserving essential information about interatomic interactions.[50-51]

Machine learning creates surrogate models that accurately represent the complex relationships captured by PELs. These surrogate models enable efficient simulations and predictions, reducing reliance on computationally expensive first-principles calculations. However, machine learning models require large, high-quality training datasets, which can be challenging to generate and curate. Understanding model predictions can be non-trivial, highlighting the need for transparent and interpretable models. Additionally, models trained on one type of PEL may not generalize well to other systems, emphasizing the importance of transferable models.[52-53]

The synergy between machine learning, experimental probes, and multiscale modeling holds significant potential for advancing our understanding and control of materials. By leveraging these approaches, new possibilities for designing and discovering materials with tailored properties for diverse applications can be unlocked.

## 7. Molecular Dynamics, Monte Carlo, and Geometry Optimization

As a theoretical concept, the PEL can be parameterized using theoretical chemistry methods, with varying levels of theory, from simple models to high-level quantum chemical methods. Molecular dynamics (MD),[54-57] Monte-Carlo (MC),[58-61] and geometry optimization represent schemes for propagating the coordinates of particles within a simulated system. MD employs

Newtonian mechanics, where particles move based on interparticle forces, with thermostat and barostat procedures controlling external conditions.

The MD methods are based on classical equations of motion for points (centers-of-mass), wherein the Hamiltonian exploits empirical force field models. Numerical integration, such as the Verlet algorithm, is needed to calculate the trajectories of motion.[62] It calculates the position of particles after a predetermined time interval. Additional routines mimic the effects of the environment on the simulated system. A thermostat algorithm, such as Berendsen's, is used as a method of maintaining temperature (total kinetic energy),

$$\frac{dT}{dt} = \frac{T_0 - T}{\tau}.$$

Here, $T_0$ is the temperature desired by the researcher, T is the instantaneous temperature calculated from the nuclear velocities, and $\tau$ is the so-called relaxation time in units of the integration time step. Subroutines are used to regulate the size of the simulation cell, such as the Berendsen barostat, to maintain constant pressure,

$$\frac{dP}{dt} = \frac{P_0 - P}{\tau_p}.$$

Here, $P_0$ is the desired pressure, P is the instantaneous pressure calculated from the interaction forces at a given integration time step, and $\tau_p$ is the given barostat time constant specified in units of the time step. In MD, researchers represent stress as a tensor quantity and relate it to the system's deformations during simulations. Pressure serves as a scalar descriptor that links microscopic quantities to macroscopic quantities.

MC utilizes a stochastic algorithm, accepting or rejecting configurations based on the Boltzmann factor, which incorporates temperature. Pressure is controlled by considering enthalpy changes. Geometry optimization methods locate the closest local minima or saddle points by following the direction of decreasing forces.

MD's advantage lies in generating real-time physical trajectories, providing direct observation of dynamic processes like diffusion, conductivity, vibrations, and conformational changes.[63] It also allows for observing non-equilibrium phenomena and tracking precise mechanisms. However, correct sampling in non-equilibrium simulations is crucial.

MC's advantage is that it does not require force computations, saving significant computer time. Unlike MD, MC can use large displacements, enabling faster movement towards the PEL's low-energy region and easier overcoming of energy barriers. This makes MC more efficient for exploring conformational space, especially for large systems and complex energy landscapes. However, this faster sampling comes at the cost of non-physical trajectories. In modern computational chemistry, MD is more frequently employed when simultaneous estimation of numerous physical and chemical properties is desired.

While both MD and MC can overcome activation energy barriers, their sampling potential is limited by the specified equilibrium temperature and system size. In unusual systems, such as viscous, glassy, or heterogeneous ones, essential configurations may remain unsampled due to limited simulation duration. Increasing the thermostat's temperature could address this but might lead to undesirable effects, such as irreversible changes in protein tertiary structure.

## 8. Global Minimum Search Problem

The global minimum, representing the lowest-energy stationary point in the PEL (Figure 4), is a ubiquitous problem across various scientific disciplines, including machine learning, optimization, and computational chemistry. Identifying the global minimum is crucial due to its correspondence to the most stable or optimal solution in a given system. However, rigorously proving the identity of a global minimum is impossible, as even when a stationary state is reported as the global minimum, absolute certainty cannot be guaranteed.[64]

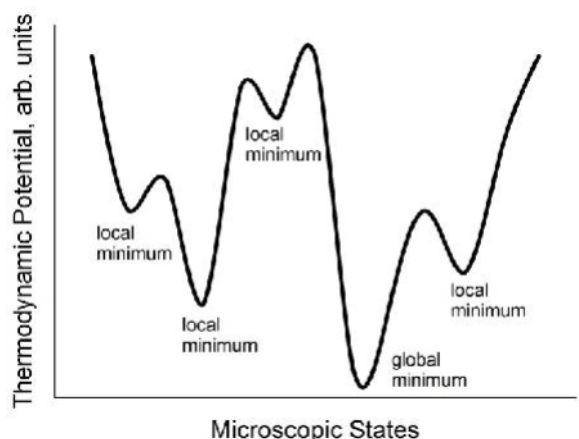

Figure 4. An exemplary potential energy surface featuring a plethora of local minima. Achievement of the global energy minimum state is challenging in numerical computer modeling but also the real-world techniques of chemical synthesis. Reproduced from Ref.[64] with permission from the Elsevier.

The need for global minimum searches arises from the complexity of real-world problems in chemistry and materials science, often characterized by high-dimensional, non-convex functions with multiple local minima. Traditional optimization algorithms, focusing on local descent, can become trapped in these local minima, yielding suboptimal solutions. Global minimum searches aim to overcome this limitation by systematically exploring the entire function landscape to identify the true global minimum, ensuring the discovery of the most optimal solution.[65-66]

Numerous methods for global minimum searches exist, each with strengths and weaknesses. These methods can be broadly categorized as deterministic or stochastic. Deterministic methods employ a systematic and exhaustive search strategy to guarantee the identification of the global minimum within a given tolerance. However, they suffer from high computational costs, especially for high-dimensional problems. Examples include branch and bound method and interval analysis. The branch and bound method divides the search space into smaller sub-regions and eliminates those that cannot contain the global minimum based on specific criteria. Interval analysis utilizes interval arithmetic to bound function values within each sub-region, enabling the elimination of regions that cannot contain the global minimum.[67-68]

Stochastic methods rely on probabilistic elements to guide the search process. While not guaranteeing the identification of the global minimum, they often provide good solutions with reasonable computational costs. Simulated annealing mimics the physical annealing process, where a system is heated and slowly cooled to reach its lowest energy state. Genetic algorithms draw inspiration from natural selection, evolving a population of candidate solutions through genetic operations. Particle swarm optimization simulates the social behavior of bird flocking or fish schooling.

The kinetic energy injection method combines MD with periodic kinetic perturbations. MD drives the system towards lower-energy states, while perturbations eliminate activation barriers that hinder the desired evolution. This method offers numerical stability and steady movement towards the global minimum.[69-71]

Challenges in global minimum searches include the absence of analytical solutions in high-dimensional space, the exponential increase in complexity with dimensionality, the presence of multiple local minima in non-convex functions, the high computational cost of many methods, and the parameter-dependent performance of stochastic methods.

Global minimum search remains crucial for identifying optimal solutions in various scientific and engineering domains. While numerous methods exist, challenges persist in addressing high dimensionality, non-convexity, and computational cost. Future research aims to develop more robust algorithms that overcome these challenges and provide reliable solutions for diverse applications.

## 9. Kinetic Energy Injection Method

As previously established, complete parameterization of the PEL, encompassing all potential energy values for all coordinates, is often unnecessary. The essential data for characterizing the PEL of new material includes the global minimum, a set of low-energy local minima, and a set of low-energy saddle points. Analyzing these critical points provides valuable insights for further investigation of specific regions to predict material properties and behaviors. This initial characterization stage focuses solely on critical points, omitting interconnections, which can be subsequently addressed using methods like elastic band calculations if needed.[40,72]

The kinetic energy injection method, a stochastic global minimum identification technique, combines MD with kinetic perturbations to overcome potential barriers (saddle points). This aims to obtain a range of lower-energy states characterizing the system. When necessary, these states can be connected via elastic band methods. The simulation starts from an initial system state defined by a complete z-matrix and evolves using MD at the temperature of interest, steadily decreasing potential energy and interatomic forces.

Periodically, the system's phase space point (q, p) in generalized coordinates is perturbed. A Maxwell-Boltzmann distribution of momenta is generated corresponding to a specified temperature and added to (q, p), leaving q unmodified.

$$f(v)d^3v = \left(\frac{m}{2\pi kT}\right)^{3/2} e^{-\frac{mv^2}{2kT}} \, d^3v$$

Alternatively, the perturbation can be constructed by multiplying all momenta by pseudorandomly generated constants, retaining the previous z-matrix and interatomic interactions but altering velocity vectors. This scheme offers numerical stability, avoids unphysical interactions, and eliminates the need for soft-core potentials. The modified momenta enable the system to overcome barriers and initiate new atomic movement directions. While affecting a limited number of atoms, other parts of the system return to their pre-perturbation positions and continue stable movement. The extra momenta is quickly removed via the coupled to the target temperature, with a coupling constant set to approximately one hundred time steps. The resulting z-matrix undergoes the steepest descent minimization to remove residual kinetic energy, representing a minimum stationary point, which is saved before the global minimum search continues from the last MD point (q, p).

The steepest descent method is a numerical method for finding a local minimum by moving along a gradient. The philosophy of such mathematical methods can be easily adapted to atomistically exact systems in chemistry and physics, in which the potential energy can be represented as a function of nuclear positions,

$$E_{pot} = \hat{H}(x_1, x_2 ... x_n).$$

Steepest descent finds the minimum of the function by propagating in the direction opposite to the gradient of the function. Once the value of the function starts to increase, it becomes possible to interpolate to an approximate minimum based on previously computed values of the function. Geometry propagation begins with the assumption $x_0$ of the minimum potential energy.

The search for the minimum begins with the assumption $x_0$ of the local minimum of the potential energy and considers a sequence $x_0, x_1, x_2 \ldots x_n$ such that.

$$E_{pot} = \hat{H}(x_1, x_2 ... x_n).$$

Thus, a uniform sequence of forces is generated until the convergence criterion $F_{conv}$ is satisfied,

$$\mathbf{F(x_0) \geq F(x_1) \geq F(x_2) \geq F_{conv}}.$$

In practical chemical calculations, a convergence criterion is often added that is related to the direct displacement of nuclei, i.e., the difference in the coordinates of each atom from the coordinate of the same atom at the previous step of geometry optimization is derived,

$$\mathbf{\Delta x = x_n(k) - x_n(k-1)}.$$

All atomic shifts must be below the convergence threshold for the complete geometry optimization to be phased out. In other words, it is considered necessary that all significant motions of the structure to be declared as a stationary point are eliminated. The kinetic energy injection method requires specifying a single parameter: the perturbation temperature. If no chemical transformation is sought, 500 K is a safe and productive option. For chemical reactions, 3,000 K provides good sampling. Since the perturbation follows a Gaussian distribution, certain atoms may receive higher kinetic energy supplements than others. Higher perturbation parameters can increase the sampling pace but may hinder movement towards the PEL's low-energy region. Other parameters follow conventional MD protocols for all-atom simulations.

Kinetic energy injection sampling can be employed with indirect perturbation magnitude control. Since most reactions occur below 3,000 K, random momenta are generated, and any momentum with kinetic energy exceeding 3,000 K is disregarded. The specific perturbation operator does not require substantiation due to the unphysical nature of the perturbations.[69]

Benchmarking suggests that approximately twenty iterations are sufficient for the system to reach the low-energy region and explore low-energy configurations, recording local minima. Minimization depends on system size and perturbation frequency. The extent of local minimizations (steepest descents) can be set to a regular number of iterations or linked to the potential energy increase following a perturbation. Internal energy, entropy, enthalpy, Helmholtz free energy, or Gibbs free energy can be used to assess the relative thermodynamic stabilities of the system.

Thermodynamic potentials can be calculated from the electronic structure of molecules.[73] The enthalpy of an ideal gas is calculated by extrapolating the energy at 0 K to the corresponding temperature,

$$H(T) = E_{elec} + E_{ZPE} + \int_0^T C_p dT.$$

Note that the enthalpy does not depend on pressure in the ideal gas. The first two terms are electron energy and zero-point energy. The integral over the heat capacity at constant pressure evaluates the energetic contribution of finite temperature. The heat capacity contains translational, rotational, vibrational, and electron parts.

The methodology employs configurational entropy to include the measure of the disorder in the simulated systems. The derived equations for translational, rotational, electronic, and vibrational components of configurational entropy are summarized below.

The translational fraction of the entropy depends on the system's molecular mass and the temperature,

$$S_{trans} = k_B \left\{ \ln \left[ \left( \frac{2\pi M k_B T}{h^2} \right)^{3/2} \frac{k_B T}{P°} \right] + \frac{5}{2} \right\}.$$

For a single atom, the rotational entropy amounts to zero. For a linear molecule, the rotational fraction of entropy amounts to

$$S_{rot} = k_B \left[ \ln \left( \frac{8\pi^2 I k_B T}{\sigma h^2} \right) + 1 \right].$$

The vibrational fraction of entropy considers all real vibrational frequencies in the system,

$$S_{vib} = k_B \sum_i^{vib\ DOF} \left[ \frac{\epsilon_i}{k_B T \left( e^{\epsilon_i / k_B T} - 1 \right)} - \ln \left( 1 - e^{-\epsilon_i / k_B T} \right) \right].$$

The electronic fraction of entropy amounts to

$$S_{elec} = k_B \ln[2 \times (spin\ multiplicity) + 1].$$

The Gibbs free energy of an ideal gas relates to the above-defined enthalpy and entropy as follows,

$$G\ (T, P) = H\ (T) - T \times S\ (T, P).$$

Depending on the thermodynamic ensemble best describing the pursued research problem, one of the thermodynamic potentials applies to rate the unraveled stationary points. The lowest-energy stationary points represent the highest interest.

The kinetic energy injection method effectively identifies low-energy local minima, in addition to the global minimum, with energetic diversity capped by the perturbation parameter. Utilizing kinetic energy perturbation instead of MC-like procedures avoids numerical problems and eliminates the need for numerous input parameters.

## 10. Sampling the PEL via a Grid

The potential energy values can be calculated at a predefined set of points on a grid that spans the relevant part of the PEL. This creates a discrete representation of the PEL, which is straightforward to implement and can provide a comprehensive view of the landscape. However, the accuracy of this method is limited by the resolution of the grid, and choosing an appropriate grid spacing for complex or unfamiliar systems can be challenging. An elementary example of a grid-based calculation would be evaluating the potential energy of a diatomic molecule at various bond lengths to map out the potential energy curve.[74-75]

Despite its limitations, the grid-based sampling method offers several advantages. It provides a standardized approach for comparing molecules with similar structures, as the same set of grid points can be used for all molecules in the comparison group. This method is also valuable for systematically investigating the effects of atomic substitutions or other structural modifications on the shape of the PEL. Moreover, grid-based sampling is numerically stable and computationally efficient, particularly for low-dimensional PELs.[76-77]

The inherent stability of the grid-based method stems from its reliance on a fixed set of predefined points in space. This eliminates the risk of divergence or instability that can sometimes occur in other sampling methods, such as those based on random sampling or dynamic trajectories. Consequently, grid-based sampling is a reliable technique for generating a preliminary understanding of the PEL's topography, especially for systems with well-defined structures or those where a comprehensive exploration of the entire landscape is not computationally feasible.[78]

## 11. The Rigid Energy Scan Method

The rigid energy scan method, a computational technique employed in chemistry and molecular physics, explores the PEL of molecules. This method systematically varies specific internal coordinates, such as bond lengths, bond angles, or dihedral angles, while keeping the remaining coordinates fixed. At each step of this systematic variation, the energy of the system is calculated, providing a discrete representation of the PEL along the chosen coordinate(s). This method is handy for initial explorations of the PEL due to its computational efficiency, requiring fewer calculations compared to more sophisticated techniques that allow for structural relaxation.[79-80]

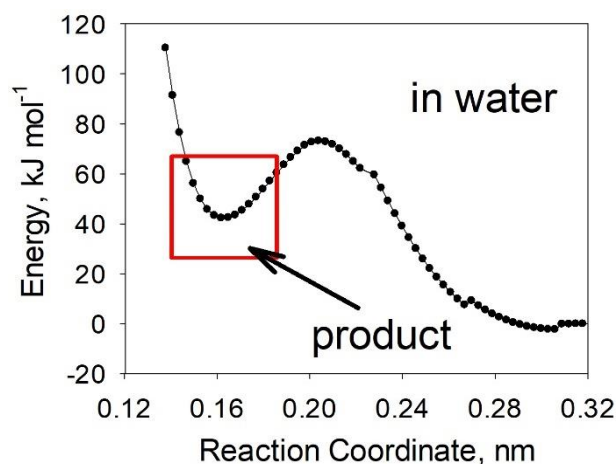

Figure 5. The reaction energetic profile for $CO_2$ reacting with the doubly-negatively-charged GQD in water. The excess charge per carbon atom of the cathode amounts to e/23. The selected reaction coordinate is a carbon (GQD) – carbon ($CO_2$) distance. Reproduced from Ref.[80] with permission from the Elsevier.

The rigid energy scan algorithm involves selecting the internal coordinate(s) to vary, defining the scan range and increment, performing single-point energy calculations at each step with fixed remaining coordinates, and finally plotting the calculated energies as a function of the scanned coordinate(s). This process yields a 1D, 2D, or higher-dimensional representation of the PEL, depending on the number of scanned coordinates.[81-83]

Despite its simplicity, the rigid energy scan method offers several advantages. Due to the reduced number of calculations, rigid scans are significantly faster than more sophisticated methods that allow for structural relaxation. This makes them ideal for preliminary investigations of the PEL, especially for large systems or when computational resources are limited. Rigid scans can help identify potential transition states by locating energy maxima along the scanned coordinate(s). This information can be used as an initial guess for subsequent transition state optimizations using more accurate methods. By fixing specific internal coordinates, the number of degrees of freedom in the system is reduced, which can significantly accelerate geometry optimizations. This is particularly beneficial for large systems where complete optimization can be computationally demanding.[84-85]

However, it is essential to acknowledge the limitations of this method. The rigid approach provides only an approximation of the true PEL because it neglects the molecule's inherent tendency to relax its geometry to minimize energy. This can lead to significant inaccuracies, especially for large changes in the scanned coordinate(s) or for highly flexible molecules. Rigid scans are not suitable for all systems or research questions. They are most effective for relatively small changes in internal coordinates and for molecules that are not overly flexible. For substantial conformational changes or highly flexible systems, more sophisticated methods that account for structural relaxation are necessary.

In summary, the rigid energy scan method is a valuable tool for initial explorations of the PEL, offering computational efficiency and facilitating the identification of potential transition states. However, it is crucial to apply this method judiciously, considering its limitations and the specific characteristics of the system under investigation.

**12. Elastic Band Methods**

Chain-of-states methods, such as the nudged elastic band (NEB) method and its derivatives, are designed to locate transition states on the PEL. These methods involve constructing a chain of intermediate system images (z-matrices) connecting the reactants and products, forming an energy reaction path. Optimization of this path yields the minimum energy path, which contains the transition state as its highest energy point.[86-87]

Chain-of-states methods require specifying two stationary points, typically the reactant and product states, for chemical reactions. The anticipated transition state geometry can also be provided as an additional starting point. A chain of preliminary points connecting the path is constructed via linear interpolation between these points, which must be in their energy-minimized states. The resulting chain, connected by virtual springs, is then optimized to satisfy convergence criteria for each image. The converged reaction path represents a sequence of points on the PEL with the lowest possible potential energies, hence the term "minimum energy path".[88-89]

The number of images significantly influences the accuracy and convergence speed of the calculation. Fewer images result in faster convergence but lower accuracy. At least four images, including reactant, product, and candidate transition states, are required. More accurate calculations may use around ten images, while larger numbers are considered inefficient due to convergence challenges.

Unlike scan-based methods, which require only the starting point, elastic band methods require knowledge of both initial and final states. However, elastic band computations offer the advantage of fixing the reaction's endpoint. In many scan calculations, guiding the system to the correct final state without introducing numerous distance restraints, which can undermine numerical stability, is challenging.[90-91]

While effective for systems with few degrees of freedom, elastic band methods are not well-suited for large, flexible systems. Moreover, they should be applied only to elementary reaction steps, as the presence of additional saddle points along the reaction path hinders convergence and prevents the generation of a minimum energy path.

## 13. The Relaxed Energy Scan Method

The relaxed energy scan method is a computational technique used to explore specific dimensions of the PEL. It involves systematically varying one or more geometrical parameters of the system, such as bond lengths, bond angles, or dihedral angles while allowing all other degrees of freedom to relax to their local or global minimum energy configurations. This approach provides valuable insights into the energy landscape of the system, revealing stable states, transition states, and minimum-energy reaction pathways. Relaxed energy scans rely on researcher-defined ranges of internal coordinates. By systematically varying these parameters and allowing the remaining degrees of freedom to relax, the method maps the energy landscape associated with a particular phenomenon. The most straightforward application of relaxed potential energy scans is to characterize chemical reactions and accompanying physical processes, such as solvation, coordination, and phase separation. If the studied process involves a significant change in potential energy, the relaxed energy scan provides beneficial microscopic insights.[92-93]

In the context of a chemical reaction, the relaxed energy scan method yields several vital results. Minimum points in the PEL correspond to kinetically stable structures, such as reactants, products, and intermediates. Transition states, representing the highest energy configurations along a reaction pathway, are saddle points connecting two minima and are crucial for understanding reaction mechanisms. The number of transition states is proportional to the system's degrees of freedom, with each transition state having its own saddle point in the PEL. Only relatively low-energy saddle points determine the chemical behavior of a molecular system. If multiple saddle points exist along a given dimension, the transformation mechanism corresponding to the lower potential energy is favored. Reaction pathways, connecting reactants to products through transition states, provide a detailed picture of how a chemical reaction or physical process unfolds.[94]

The relaxed energy scan procedure typically involves defining the initial state as a z-matrix, selecting the PEL dimension to be scanned (often referred to as the reaction coordinate), choosing the quantum chemical method or force field model for energy calculations, and propagating the system's geometry along the selected dimension(s) to connect the initial and final states through saddle points and intermediate minima. While the reaction coordinate is rigidly fixed, all other coordinates relax to adapt to the changes. The spatial step for propagating the reaction coordinate is typically 1-10 pm but can be increased to 100 pm for preliminary scans. Larger steps are not recommended when focusing on chemical phenomena. Specific PEL dimensions are naturally expressed as planar or dihedral angles, allowing for scans of functional group rotations. The resulting PEL parameterizations can be used to develop empirical models for simulating large, flexible molecules like nucleic acids, polypeptides, and polysaccharides via MD and MC methods. Finally, a plot of the determined potential energies as a function of the scanned geometrical parameters visualizes the parameterized PEL fraction with its stationary points.

One valuable product of relaxed potential energy scans is the energies of saddle points, which characterize activation barriers. This allows for reliable predictions of reaction pathways favored in real-world syntheses, enabling significant portions of new syntheses to be developed and rationalized in silico. The importance of adding the zero-point energy correction should be noted.

The relaxed energy scan method is a versatile technique applicable to various molecular systems and processes. It is handy for studying conformational changes, chemical reactions, and intermolecular interactions. Exploring different molecular conformations and their relative energies is crucial in biophysical research, particularly for protein folding. Mapping reaction pathways and identifying transition states aids in catalysis development and rationalizing organic and hybrid syntheses. Investigating intermolecular interactions, such as hydrogen bonding and van der Waals forces, is essential in various materials science fields, from optimizing electrolyte solutions to understanding gas adsorption.

The relaxed energy scan method can be a valuable tool in developing new functional materials. By exploring the PEL of different materials, researchers can identify stable structures, predict their properties, and design materials with desired characteristics. Phase transitions can be investigated by scanning the PEL as a function of temperature or pressure, allowing for the identification of transition conditions and prediction of phase properties. Defects in materials can also be studied by scanning the PEL around a defect to understand its stability and influence on material behavior. Additionally, the method can be used to study molecular adsorption on surfaces, which is vital for understanding catalytic processes. By scanning the PEL of a molecule on a surface, researchers can identify stable adsorption sites and minimum-energy reaction pathways.

The PEL's curvature changes from concave to convex in valley ridge inflection points. The latter are not necessarily saddle points. They can have zero or more imaginary frequencies. While less critical in chemistry than saddle points, they can indicate bifurcations in the PEL, leading to different products depending on reaction conditions. The relaxed energy scan method can locate these points. The Hessian matrix, containing the second derivatives of potential energy with respect to PEL coordinates, is analyzed to confirm their identity. Diagonalization of the Hessian provides its eigenvalues at a given PEL point. A valley-ridge inflection point has one zero eigenvalues, and the remaining eigenvalues have the same sign. This combination of eigenvalues indicates a change in curvature along one direction. In some cases, inflection points can be visually identified by analyzing energy evolution along a PEL dimension. Researchers can locate these critical points by systematically varying geometrical parameters. Next, one must analyze vibrational frequencies while gaining a deeper understanding of the energy landscape.

## 14. Metadynamics-Based Sampling of the PEL

Metadynamics, a biased MD method, enhances the sampling of rare events, which are ubiquitous in microscopic phenomena. This enhancement is achieved by introducing a bias potential to the model Hamiltonian, effectively modifying the PEL and increasing the likelihood of observing these rare events. The bias potential discourages the system from revisiting previously sampled regions of the PEL, compelling it to explore new areas. Successful application of metadynamics necessitates a thorough understanding of the simulated system, its sampling challenges, and the overall topography of the PEL. A judiciously applied metadynamics simulation can efficiently explore rare events, overcome energy barriers, and provide insights into complex processes such as protein folding and phase transitions.[95-96]

However, metadynamics has limitations. The selection of appropriate bias potential parameters requires careful consideration, and the available information about the system may be insufficient for this task. Additionally, metadynamics simulations can be computationally demanding. Despite these limitations, metadynamics is frequently employed in biophysical research, where sampling challenges cannot be readily addressed by simply elevating the temperature or employing replica exchange methods. Although metadynamics facilitates the exploration of numerous microscopic states, it is unsuitable for deriving equilibrium transport properties.[97-98]

Metadynamics is particularly well-suited for investigating transitions between different crystal structures, provided a reliable Hamiltonian can be selected. By introducing a bias potential that discourages the system from remaining in its initial crystalline state, thermal motion drives the system towards higher potential energies, thereby enhancing the sampling of the PEL and facilitating the observation of phase transitions.[99-100]

## 15. Case Study: Self-Healing in Dielectric Capacitors

Dielectric capacitors are essential components in electronic devices, storing electrical energy and filtering signals. However, they are susceptible to electrical breakdown caused by structural defects, poor electrode/insulator contact, or overvoltage. Such breakdowns can lead to device failure, energy loss, and capacitor damage. Self-healing is a valuable property exhibited by certain dielectric materials, enabling them to recover from breakdowns and continue operating. This case study examines the mechanisms of self-healing within the framework of the PEL.[63]

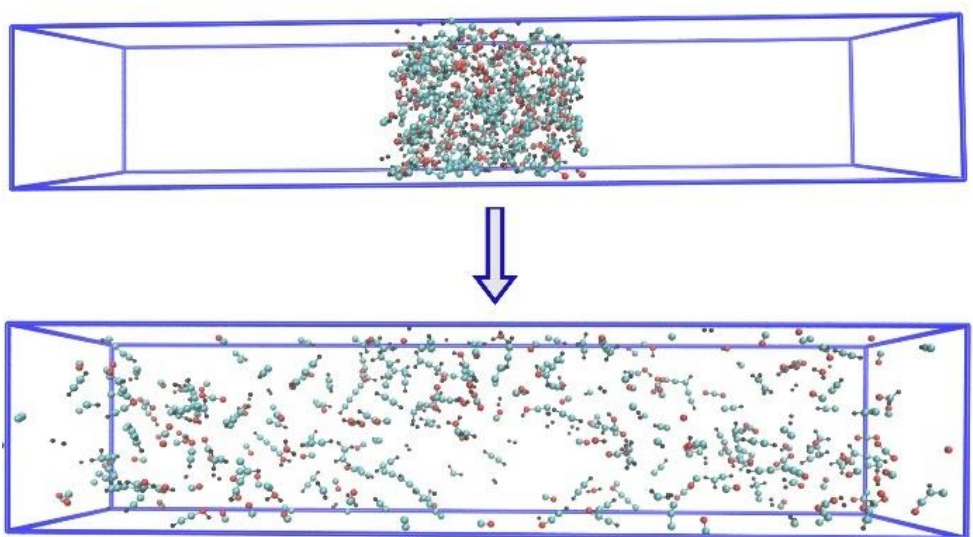

Figure 6. The exemplary simulation cell containing PET (1 chain, 42 monomers) at the beginning (top) and at the end (bottom) of the RMD simulations. The dimensions of the simulation cell amount to $2.60 \times 2.28 \times 10.4$ nm$^3$. The oxygen atoms are red, the carbon atoms are cyan, and the hydrogen atoms are black. Reproduced from Ref.[63] with permission from the Royal Society of Chemistry.

Electrical breakdown in a dielectric material creates a conductive path between the electrodes. The high current density at the breakdown site generates intense heat, vaporizing the electrode material and surrounding insulator, which is typically a polymeric material. The rapid expansion of the vaporized material creates pressure that pushes the molten electrode material away from the breakdown site, isolating the fault and interrupting the current. The vaporized material dissipates or deposits on the capacitor electrodes, leaving an insulated gap where the breakdown occurs. This process restores the capacitor's functionality, although with a slightly reduced capacitance.

Metalized film capacitors utilize thin metal layers deposited on a polymer film as electrodes. This design facilitates effective vaporization and fault isolation. Zinc, aluminum, or zinc-aluminum electrodes are usually used as plates.[101]

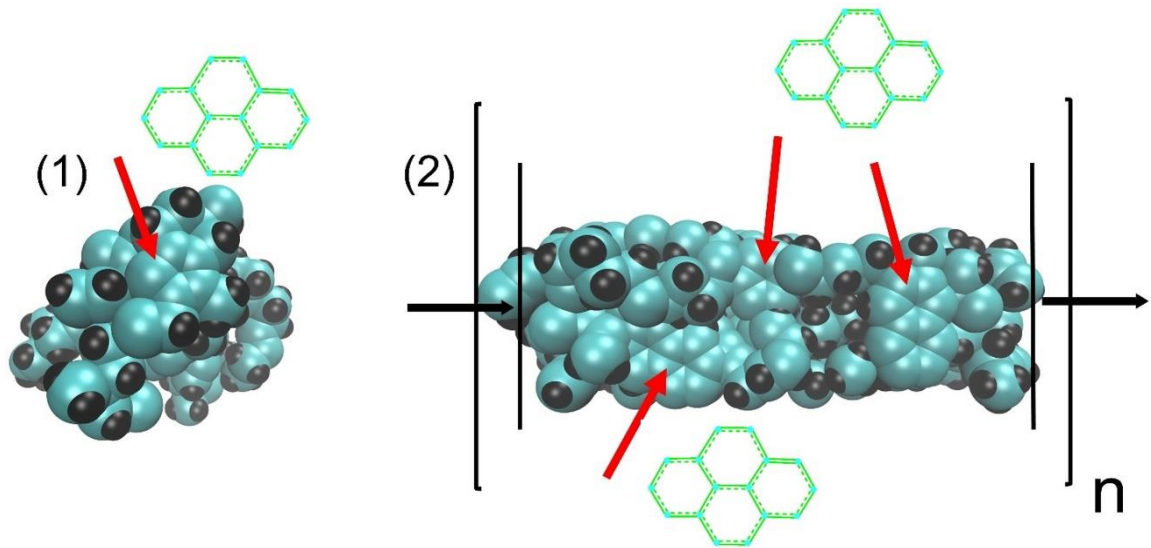

Figure 7. The optimized molecular geometries of the carbon quantum dots were obtained out of the PP polymer: (1) non-periodic spherical prototype and (2) after applying the periodicity along one of the Cartesian coordinate directions. The red arrows point to the hydrogen-free conjugated carbon rings. Reproduced from Ref.[63] with permission from the Royal Society of Chemistry.

Self-healing is crucial in various applications of dielectric capacitors. In high-power applications, self-healing capacitors enhance reliability and prevent catastrophic failures. Metalized film capacitors are used in receiving and transmitting radio equipment, electrical measuring equipment, lasers, and X-ray units. In consumer electronics, they improve the lifespan and reliability of portable devices.

Research efforts are focused on enhancing the speed and efficiency of self-healing while minimizing the impact on capacitor performance. The number of self-healing events a capacitor can withstand is finite, and ongoing research aims to increase this capacity. Understanding the microscopic basis of electrical breakdown mechanisms is essential, as self-healing mechanisms vary for different insulating materials. A deeper understanding of these mechanisms is crucial for optimizing self-healing parameters and developing more robust dielectric polymers.

## 16. Case Study: Developing Innovative Electrolyte Solutions

Electrolytes, substances that conduct electricity via ionic conduction, are crucial in various applications, including batteries, fuel cells, electrochemical sensors, and medical devices. At room temperature, most electrolytes exist in a solid or highly viscous liquid state, necessitating dissolution in a solvent to form a functional electrolyte solution. Developing innovative electrolyte solutions is paramount for enhancing the performance, safety, and sustainability of these technologies. This case study examines the optimization of electrolytic systems using the PEL framework to guide the development of next-generation electrolytes.[102-104]

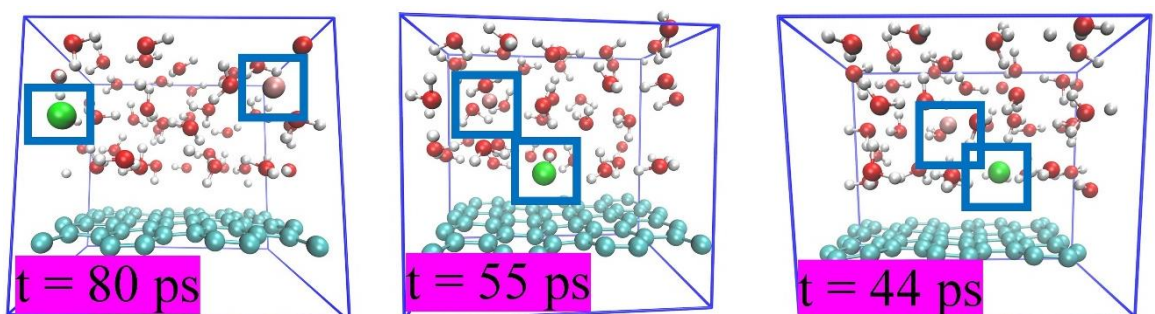

Figure 8. Molecular geometries of the ion pairs at the graphene surface in water taken during the equilibrium ab initio molecular dynamics simulations at 350 K. The chosen time frames are arbitrary. The cation is pale pink, the anion (chloride) is green, the oxygen atoms are red, and the hydrogen atoms are white. The immediate locations of the ions are highlighted by dark blue rectangles. Reproduced from Ref.[102] with permission from the Elsevier.

The performance of electrolytes is intrinsically linked to their ionic conductivity. High conductivity ensures efficient charge transport in electrochemical devices. Traditional liquid electrolytes often exhibit limitations at low temperatures or high concentrations. The incorporation of solvents and cosolvents, such as room-temperature ionic liquids, aims to enhance the capabilities of energy storage devices at low temperatures. Additionally, addressing the high shear viscosity of electrolytes, which hinders ionic mobility, is crucial for materials design.[36,40,105]

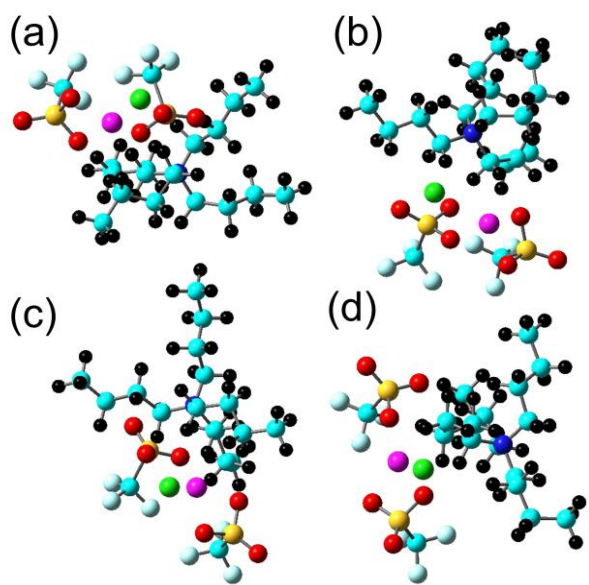

Figure 9. (a) The global minimum geometry of the [Mg][OTf]$_2$ + [TBA][Cl] system with a potential energy of -3245 kJ/mol; (b) the local minimum geometry with a potential energy of -3084 kJ/mol; (c) the local minimum with a potential energy of -3242 kJ/mol; (d) the local minimum with a potential energy of -3234 kJ/mol. The carbon atoms are cyan, the hydrogen atoms are black, the nitrogen atoms are blue, the fluorine atoms are light blue, the oxygen atoms are red, the sulfur atoms are yellow, the chloride anions are green, and the magnesium cations are violet. Reproduced from Ref.[40] with permission from the Elsevier.

Electrolyte stability over a wide voltage range is essential to prevent undesirable side reactions and degradation, compromising device performance and safety. Flammability, toxicity, and leakage are major safety concerns associated with conventional liquid electrolytes, particularly in high-energy applications like electric vehicles. Furthermore, many electrolyte materials raise environmental problems due to their toxicity, resource depletion, or complex disposal processes. These challenges can be addressed through materials optimization strategies guided by the PEL concept.[69,106-108]

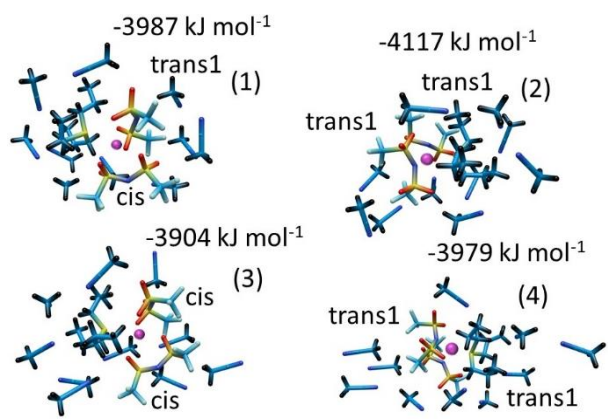

Figure 10. The global (top) and local (bottom) minima molecular configurations in the [Li][S$_{333}$][TFSI]$_2$×10 ACN (left) and [Na][S$_{333}$][TFSI]$_2$×10 ACN (right) systems. Reproduced from Ref.[69] with permission from the Elsevier.

Several promising research directions are being pursued nowadays. Replacing liquid electrolytes with solid-state materials offers improved safety and stability. Research focuses on developing solid-state electrolytes with high ionic conductivity, wide electrochemical windows, and robust mechanical properties. Systematic research in this area necessitates exploring the corresponding PELs to compare various candidate materials and their combinations. Solid-state electrolytes enable the development of safer and more energy-dense batteries for electric vehicles and grid storage.

Ionic liquids, salts composed of asymmetric ions that remain liquid at room temperature, possess desirable properties such as moderate ionic conductivity, wide electrochemical windows, and low flammability. However, their high viscosity and cost can be limitations. The kinetic injection method offers valuable theoretical support for addressing these challenges and generating practical recommendations for electrolyte design.[109-110]

Polymeric electrolytes, formed by combining polymers with salts, yield flexible and lightweight electrolytes with excellent mechanical properties. Current challenges include achieving high ionic conductivity and wide electrochemical windows.[35, 111]

Redox-active electrolytes actively participate in electrochemical reactions, increasing energy density and enabling new battery chemistries. Research focuses on understanding and controlling their redox properties. Computational tools like MD and DFT simulations predict cation-anion coupling in multicomponent systems, competing solvation effects, and interactions of electrolytes with electrodes. Understanding these microscopic phenomena through in silico PEL sampling provides insights into electrolyte properties and guides the design of new materials.

New battery designs are constantly emerging. Lithium-sulfur batteries offer higher theoretical energy density than lithium-ion batteries, but efficient constructions remain under development. Sodium-ion batteries are more cost-effective and environmentally friendly, but their longevity is unlikely to surpass that of lithium-ion batteries. PEL-driven optimization can enhance existing technologies and guide the development of new electrolyte formulations for these advanced battery systems. Fuel cells also require efficient and durable electrolytes, and developing such electrolytes is crucial for advancing fuel cell technologies for clean energy applications.[103, 112]

Developing innovative electrolyte solutions, potentially through fine-tuning existing systems, is essential for advancing electrochemical technologies. Researchers can create safer, more efficient, and sustainable electrolytes for various applications by overcoming challenges and exploring new approaches. PEL sampling is crucial in this endeavor by providing a framework for understanding and tailoring multi-component systems with complex non-covalent interactions. Comparing aggregate potential energies within the PEL framework offers a unified methodology for guiding ongoing research and development efforts.

**17. Case Study: Greenhouse Gas Scavenging**

Greenhouse gas emissions, primarily carbon dioxide ($CO_2$), methane ($CH_4$), and nitrous oxide ($N_2O$) are the main drivers of climate change. Scavenging these gases from the atmosphere or industrial sources is crucial for mitigating their environmental impact. However, successful adsorption and absorption of gases require a precise understanding of the interactions between the sorbent and sorbate. This case study discusses the challenges in greenhouse gas scavenging and the role of PEL sampling in developing robust sorbents.[113-116]

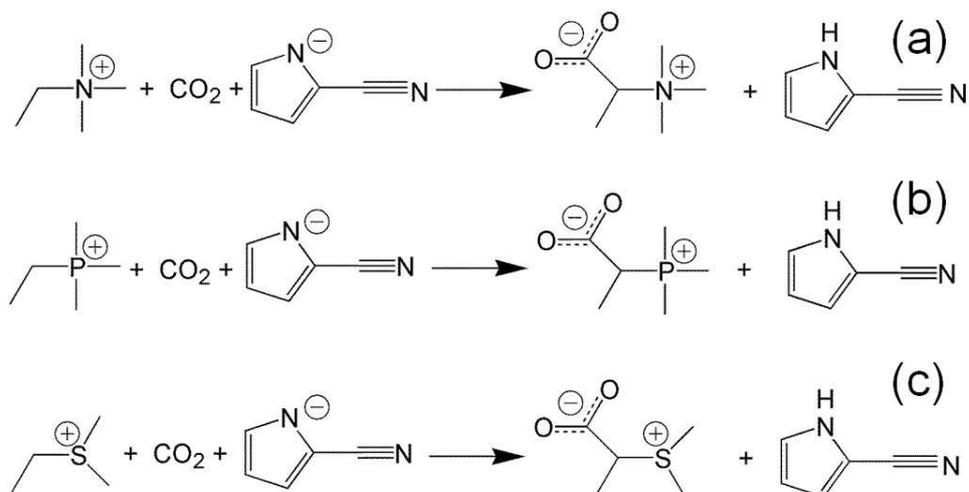

Figure 11. The schemes summarizing $CO_2$ grafting to α-carbon atoms of the investigated acyclic cations: (a) $[N_{2111}]^+$; (b) $[P_{2111}]^+$; (c) $[S_{211}]^+$. Note that all chemisorption reactions are appropriately balanced. Reproduced from Ref.[116] with permission from the Royal Society of Chemistry.

The low concentration of greenhouse gases in the atmosphere makes their capture challenging and energy-intensive. Efficient separation from other gases requires selective capture methods. Developing scalable technologies for capturing significant amounts of $CO_2$ and $CH_4$ is crucial. Energy consumption and associated economic costs are essential factors in evaluating new technologies. Many current scavenging processes require considerable energy input, potentially offsetting environmental benefits. Costs associated with gas desorption for sorbent renewal also pose implementation barriers.[117-119]

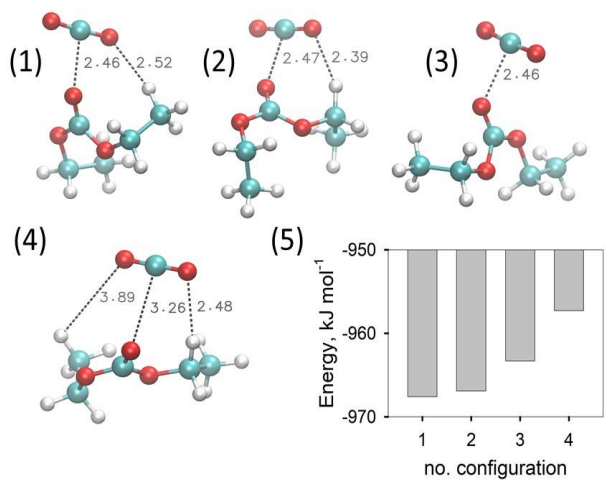

Figure 12. (1-4) local minimum energy configurations in the system containing one D2C molecule and one $CO_2$ molecule. The chosen closest-approach atom-atom distances are provided in Å, dashed lines. (5) The potential energies of the recorded stationary points. The provided potential energies were obtained in the framework of the PM7 Hamiltonian. The carbon atoms are cyan, the oxygen atoms are red, and the hydrogen atoms are white. Reproduced from Ref.[117] with permission from the Royal Society of Chemistry.

Direct air capture is currently energy-intensive and expensive when using chemical processes to capture $CO_2$ from ambient air. Capturing $CO_2$ from concentrated sources like power plants is more efficient. Relevant technologies include post-combustion capture, pre-combustion capture, and oxy-fuel combustion.[120-121]

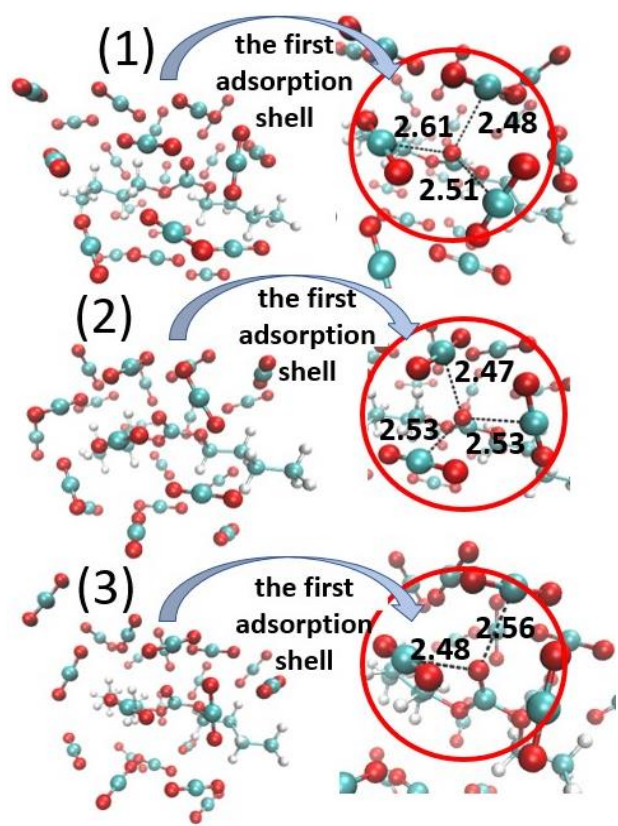

Figure 13. (1) Global and (2-3) local minima in the system containing one dibutyl carbonate molecule and 21 $CO_2$ molecules. Note that the simulated system contains many more minima, whereas we chose only a few of the most representative ones for ease of representation. The chosen closest-approach atom-atom distances are provided in Å, dashed lines. The carbon atoms are cyan, the oxygen atoms are red, and the hydrogen atoms are white. Reproduced from Ref.[117] with permission from the Royal Society of Chemistry.

Biological scavenging utilizes natural processes like photosynthesis by plants, algae, or engineered microbes to absorb $CO_2$. Afforestation and reforestation are examples of this approach. Enhancing the ocean's natural $CO_2$ absorption capacity through geoengineering is also being investigated. Specific geoengineering methods include ocean fertilization and artificial upwelling. Ocean fertilization involves introducing nutrients like iron and urea to stimulate phytoplankton growth and $CO_2$ uptake. Artificial upwelling mimics natural processes that bring nutrient-rich water to the surface. However, both methods have potential drawbacks, including energy requirements, cost, and ecological consequences.

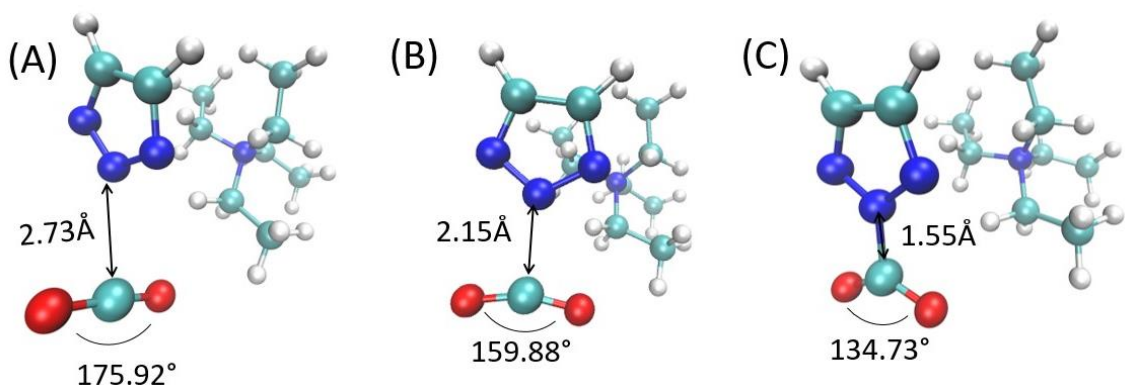

Figure 14. The representative molecular geometries were recorded in water during the guided $CO_2$ chemisorption by tetraethylammonium 1,2,3-triazolide. Reproduced from Ref.[79] with permission from the Elsevier.

Captured $CO_2$ can be stored or utilized. Carbon capture and storage involves compressing and injecting $CO_2$ into geological formations. Carbon capture and utilization uses $CO_2$ as a feedstock for producing valuable products. Biomass energy production can be combined with carbon capture and storage to minimize harmful emissions.

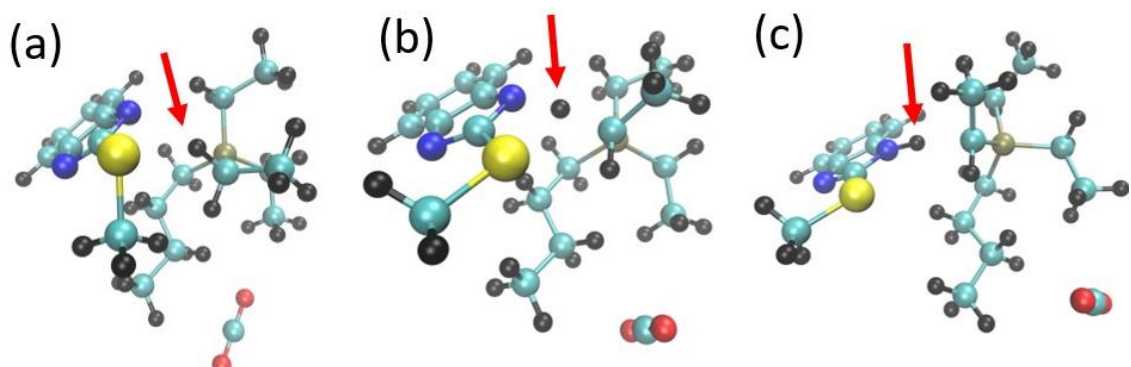

Figure 15. The representative molecular geometries for the deprotonation reaction (ylide formation) of the triethyl-butyl phosphonium cation in triethyl-butyl phosphonium 2-methylthio-benzimidazolide. (a) Energy minimum for the initial state. (b) Transition state geometry. (c) Energy minimum after ylide formation. The oxygen atoms are red, the hydrogen atoms are black, the carbon atoms are cyan, the sulfur atoms are yellow, and the phosphorus atoms are grey. The red arrows point to the exchanged hydrogen atom. Reproduced from Ref.[81] with permission from the Elsevier.

Developing new materials for efficient $CO_2$ capture, such as metal-organic frameworks and advanced membranes, can increase sorbent capacity. PEL-driven materials design may reduce preparation costs. However, the ultimate solution to the $CO_2$ problem lies in finding truly green and efficient energy sources and phasing out oxidation-based energy production.[113,122-123]

Greenhouse gas scavenging is critical for mitigating climate change. While technological challenges persist, ongoing research and development lead to more efficient and cost-effective technologies. Computational methods for PEL parameterization provide a framework for systematically enhancing functional materials.

**18. Conclusions and Final Remarks**

This review paper provides a comprehensive overview of the PEL concept. The evolution of the PEL, from the foundational work of Gibbs to its contemporary applications, is traced, with specific examples highlighting its role in advancing materials design. Additionally, the review explores modern computational methods used to navigate the PEL at various levels of theoretical complexity. The case studies illustrate the PEL's utility as a framework for addressing specific challenges related to self-healing in dielectric capacitors, electrolytic solutions, and greenhouse gas capture.

The PEL serves as a guiding principle, offering a unified framework for understanding the structure and dynamics of materials. It provides a multidimensional map of the energy states accessible to the interacting components within a material.

Analyzing the shape of this landscape allows researchers to predict material behavior under different conditions, such as variations in temperature or pressure. The PEL elucidates how atomic arrangements and interactions influence fundamental material properties, guiding the design of novel materials with tailored properties.

Furthermore, the PEL concept enables the control of transitions between different states of matter, including crystallization, melting, evaporation, and glass formation. This capability has significant implications for developing new processing techniques and manipulating material behavior. Although often used implicitly, the PEL framework is pervasive in materials science.

It is important to note that the PEL concept is not a direct research method for materials design. Instead, it provides a language for discussing materials science in terms of atomic coordinates and potential energy values. The PEL facilitates a more rational design of materials with customized properties, leading to the development of advanced materials for energy storage, electronics, electrical technologies, and gas sorption. Advances in PEL research enhance the accuracy of predicting material behavior under diverse conditions, enabling the development of more durable and reliable materials. Researchers can control properties, phase transitions, chemical reactions, catalytic applications, and other material aspects by deliberately manipulating the PEL.

**Credit Author Statement**

Author 1 : Conceptualization; Methodology Development; Software development; Validation; Formal analysis; Investigation; Resources; Data Curation; Writing - Original Draft; Writing - Review & Editing; Visualization Preparation; Supervision; Project administration; Funding acquisition.

Author 2 : Conceptualization; Methodology Development; Software development; Validation; Formal analysis; Investigation; Resources; Data Curation; Writing - Original Draft; Writing - Review & Editing; Visualization Preparation; Supervision; Project administration; Funding acquisition.

**Conflict of interest**

The authors hereby declare no financial interests and professional connections that might bias the interpretations of the obtained results.

**Acknowledgments**

The research was funded by the Ministry of Science and Higher Education of the Russian Federation under the strategic academic leadership program "Priority 2030" (Agreement 075-15-2024-201 dated 06.02.2024).